# Superstripes : superconductivity in complex granular striped matter

Antonio Bianconi[1], Davide Innocenti[1], Gaetano Campi[2]

[1]*RICMASS Rome International Center for Materials Science, Superstripes Via dei Sabelli 119A, 00185 Roma, Italy.*
[2]*Institute of Crystallography, CNR, Via Salaria, I-00015 Monterotondo, Rome, Italy.*

E-mail: antonio.bianconi@ricmass.eu

Abstract

The role of *phase separation and superstripes in high temperature superconductors and related materials has been discussed at the meeting in* Erice, Italy, July 11–17, 2012. The focus was on the point where we are after 25 years of research on the key role of complex structural, electronic, and magnetic matter in the mechanism of high temperature superconductivity.

The present special issue reports key talks presented at the third superstripes 2012 meeting on *"Phase Separation and superstripes in high temperature superconductors and related materials"* Erice, Italy, July 11–17, 2012 chaired by Antonio Bianconi and Giorgio Benedek. The first superstripes *meeting has focused on "'FeAs High $T_c$ Superconducting Multilayers and Related Phenomena'. It was held in* Rome, Italy 9–13 December 2008 and the second superstripes 2010 meeting focused on *Quantum Phenomena in Complex Matter*. *It was held in* Erice, Italy, July 19–25, 2010. The topic of the meeting was the role of phase separation in the mechanism that allows the macroscopic superconducting quantum condensate to resist to the de-coherence attacks of temperature giving high temperature superconductivity (HTS). The series of superstripes conferences focus on i) the complex structure of the lattice and ii) the complex electronic phase of the materials where this unique phenomenon emerges. The issue is whether electron-electron interactions can lead to intriguing pairing ground states and especially whether periodicity in the underlying lattice and the possible occurrence of complex structures can enhance the stability of the pairing ground states. In fact, the effective interactions can be appreciably modified in systems with discrete translational symmetry and especially two - or more – Fermi surfaces with different symmetry and eventually with electron and hole characters.

The 1957 BCS theory considers isotropic s-wave pairing [1,2]. The superconducting phase can be simply understood from the viewpoint of spontaneous symmetry breaking. In early seventies Leggett [3] proposed p-wave pairing to describe the properties of $^3$He. The Nabu-Jona-Lasinio mechanism [4,5] of spontaneous broken symmetry in subatomic physics is the basis for both the strong interaction's chiral symmetry and the exchange mechanism of Higgs boson for the formation of massive elementary particles. To describe the properties of atomic nuclei Iachello proposed a combination of s- and d-wave pairing [6,7]. Moreover, complex superconductivity occurs in neutron stars [8,9]. While the BCS theory for standard metals considers an infinite electronic system with high Fermi energy, for an electron gas confined in a single slab [10] or in a superlattice of slabs [11], in a single wire [12] or in a superlattice of wires [13] quantum size effects produce irreducible different electronic components at the Fermi level with odd and even symmetries; this give rise to multiple condensates where standard single band BCS approximations breakdown. In these systems where the Fermi level is tuned to a 2.5 Lifshitz transition [14] i.e., a metal-to-metal transition from a N-band metal to a (N+1)-band metal or vice versa, the superconducting phase cannot be reduced to an effective single-band superconductor. Thus the theory of multi-band BCS superconductivity [15] becomes the key theoretical scheme to describe the new physics due to key role of configuration interactions between different condensates driven by exchange interactions, also where one of the condensate is bosonic-like as shown in Fig. 1 [16]. In this regime both the exchange of lattice fluctuations and coulomb magnetic interactions become relevant as indicated by an anomalous isotope effect with an isotope coefficient strongly dependent on the chemical potential [17]. Crossing the band edge, the shape resonance in the superconducting gaps is shown to be a particular realization of Fano resonance between closed and open scattering channels [18-21].

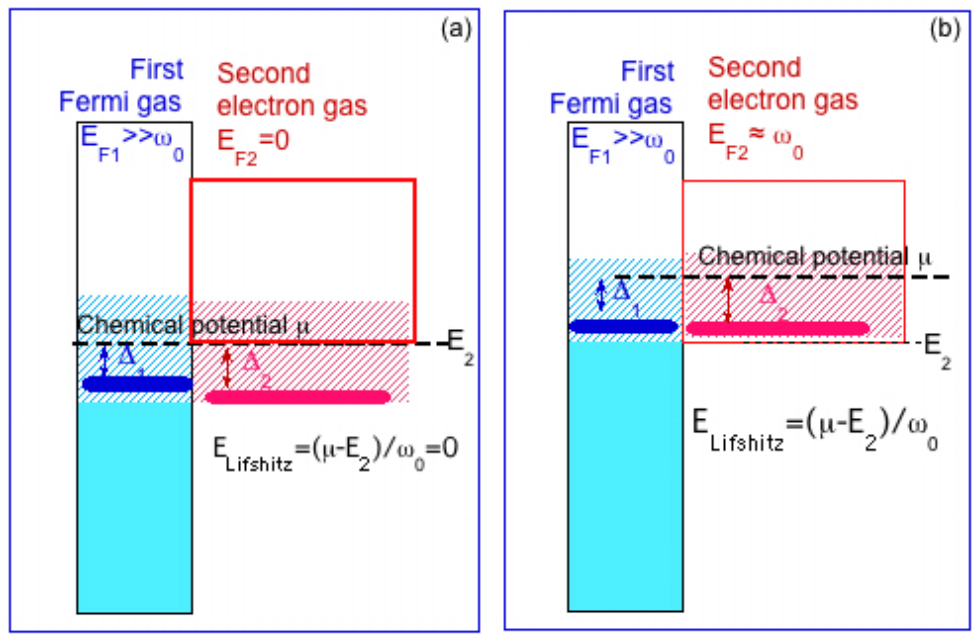

**Figure 1**: *Metals made of two bands with different symmetry where the chemical potential is tuned at a metal (single band) to a metal (two bands) transition called 2.5 Lifshitz transition. The two different electronic components have different symmetry even and odd due to quantum size effect, or π and σ states in boron or graphene superlattices. Panel a) shows the "2.5 Lifshitz transition for the appearing of a new Fermi surface spot" in the normal phase and in the superconducting phase: the formation a fermionic condensate with gap ($\Delta_1$) in the large Fermi surface and a bosonic condensate with gap ($\Delta_2$) in the second set of states below the bottom of the second band; Panel (b) shows the case where in the normal phase the second components is in the polaronic regime where the Fermi energy separation from the bottom of the band is of the order of the electron-electron interaction $\omega_0$.*

It was known that the lattice of cuprates is a particular heterostructure at *atomic* limit formed by the active superconducting $CuO_2$ atomic layers. In 1993 it was first proposed [22,23] that the architecture shown in Fig. 2, superlattices of superconducting atomic layers intercalated by spacer layers, is an essential ingredient for materials design of high temperature superconductors.

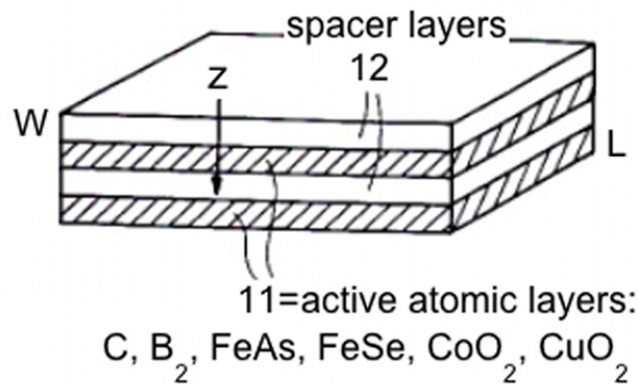

**Figure 2**: *The lattice architecture of heterostructures at atomic limit showing high temperature superconductivity. The superlattice of quantum wells is made of stacks of active 2D atomic metallic layers (11) : $CuO_2$ in cuprates, graphene in intercalated graphite, exagonal boron layers in diborides, $CoO_2$ layers in cobaltates, fluorite FeAs layer in iron based pnictides, or FeSe layers in selenides. The active layers are separated by spacer layers 12.*

This prediction has been first confirmed in 2001 by diborides, later by intercalated graphite, in 2008 by iron pnictides and in 2010 by selenides where the active layers are $B_2$, C, FeAs, and FeSe, respectively. This lattice architecture realized by graphene-like layers is nowadays the focus for new nanoscale functional materials for electronics at atomic level.

The high temperature superconductivity emerges in the active layers by driving the chemical potential in the proximity of a 2.5 Lifshitz transition where multiband superconductivity appears. It was first confirmed in $Al_xMg_{1-x}B_2$ [24,25,11] and recently clearly observed in cuprates near 1/8 doping [26], in electron doped iron based superconductors Ba(Fe$_{1-x}$Co$_x$)$_2$As$_2$ [27] and (Fe$_{1-x}$Ni$_x$)$_2$As$_2$ [28] as predicted by Innocenti et al [29].

In the materials with the critical temperature above liquid nitrogen the point group invariant is not the D$_{4h}$ symmetry. YBa$_2$Cu$_3$O$_{7-y}$ shows the C$_{2v}$ point group symmetry for oxygen sites hosting the itinerant electronic holes, and D$_{2h}$ and C$_{2v}$ point group site symmetries for Cu(2) (in the plane) and Cu1 (in the chains) site, respectively. In Bi$_2$Sr$_2$CaCu$_2$O$_{8+y}$ with higher temperature the atoms of the superconducting CuO$_2$ planes are on C$_{2v}$ and C$_2$ sites. Moreover intricate incommensurate structural modulations with bond modulations in the CuO$_2$ and in the spacer layers are observed. The atomic planes are instable toward i) local lattice distortions [30,31] detected by XANES [32-34] and EXAFS [35] ii) ripples [36] detected by resonant x-ray diffraction iii) charge localization in oxygen versus copper site viewed using x-ray absorption edge spectroscopy [37] iv) non homogenous distribution of defects with puddle formation by self organization of oxygen interstitials in the spacer layers [38,39,40] that control electronic states at the Fermi energy [41]. Looking at the mesoscopic scale of 5-100 nanometers a complex granular structure of network of nano-puddles is observed ( see Fig. 3) made of an arrested phase separation of striped nanoscale puddles showing a superstripes scenario [42-48] providing support for theoretical proposals of phase separation [49-53]; therefore, the scenario of granular networks of superconducting puddles has been proposed where the maximum T$_c$ occurs where percolation occurs [54] and recently theories for networks of Josephson junctions and scale free networks have been proposed to rise the critical temperature [55-59].

The superstripes conferences follow the conferences on *Stripes and related phenomena* [60] and Erice workshop on *Symmetry and heterogeneity in high temperature superconductors* [61]. The name superstripes was introduced in 2000 at the international conference on "Stripes and High T$_c$ Superconductivity" held in Rome to describe a particular phase of matter where a broken lattice symmetry - appearing at a transition from a phase with higher dimensionality N (3D or 2D) to a phase with lower dimensionality N-1 (2D or 1D) - does not compete with the superconducting or superfluid phase but on the contrary it increases the normal to superconducting transition temperature with the possible emergence of high-temperature superconductivity.

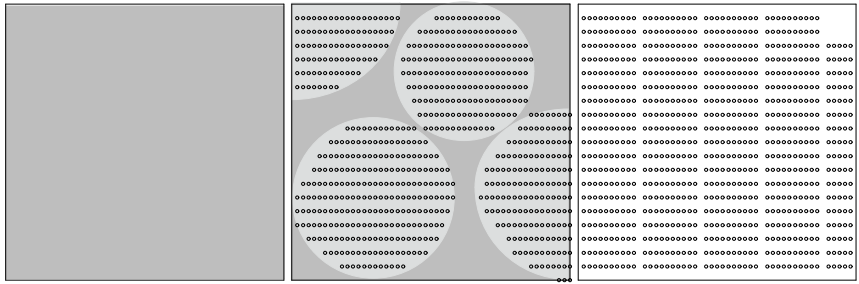

**Figure 3**: *The metallic 2D active layers in the superconducting heterostructures at atomic limit show a "superstripes" scenario made of bubbles of striped metallic matter in a different matrix (b) that is a metastable phase between the homogeneous metallic phase (a) and the striped crystalline matter (c) controlled both by i) lattice via dopants self organization in the spacer layers ii) the lattice misfit strain between the different layers and iii) the electronic phase separation in the electron gas due to proximity to a 2.5 Lifshitz transition and/or coulomb interactions in the Hubbard multi band scenario.*

Therefore, in the broken symmetry of superstripes phase the structural modulation coexists and favors high temperature superconductivity. The name *superstripes* was chosen to indicate the key difference with the *stripes* scenario where the phase transition from a phase with higher dimensionality N (like a 2D electron gas) to the phase with broken symmetry and lower dimensionality N-1 (like a quasi 1D striped fluid) competes and suppresses the transition temperature to the superfluid phase and favors modulated striped magnetic ordering. Superstripes is a generic name for a phase with spatial broken symmetry that favors the onset of superconducting or superfluid quantum order. This scenario emerged in the nineties when particular doped oxide hetero-structures at the atomic limit with a broken spatial symmetry have been found to favor superconductivity. Usually a broken spatial symmetry was expected to compete and suppress always the superconducting order. Superstripes are type of supersolids [62] i.e., phases displaying simultaneously different types of order. In a crystalline solid, elementary constituents (i.e., atoms or molecules) are spatially arranged in an orderly fashion, each occupying one of a discrete set of well-defined sites of a three-dimensional periodic lattice. For the particles confined to move on discrete lattices the spontaneous breaking of translation invariance is defined with reference to the discrete translation symmetry of the Hamiltonian with opening of gaps in the one-particle spectrum. In superstripes the presence of the superconducting order of the electronic fermion gas, referred as off-diagonal long range order (ODLRO), coexists with broken symmetry of the lattice below a critical temperature. In superstripes the lattice heterostructure promotes the formation of the multiple superconducting gaps, i.e. different condensates with different order parameters of the off diagonal superconducting order. Therefore, superstripes are a particular case of multiband superconductors or multigap superconductors where exchange terms related with configuration interaction between different pairing channels show interferences that can produce both negative interference effects like Fano antiresonances, and positive effects like Fano resonances, called shape resonances in the superconducting gaps. The topic of the Erice 2012 meeting was on phase separation in high temperature superconductors; in fact the open problem is the role of complex superconducting networks on superstripes puddles where the gaps are not only different in different portions of the k-space but also in different portions of the real space.

**References**

1. Schrieffer JR 1964 *Theory of Superconductivity*, (for exchange-like pair transfer mechanism see 300 pp.)
2. Schrieffer JR 1992 *Phys. Today* **45**, 46.
3. Leggett, A. J. 1975 *Reviews of Modern Physics* **47**, 331-414
4. Nambu Y. and Jona-Lasinio, G. 1961 *Physical Review* **122**, 345
5. Nambu Y. and Jona-Lasinio, G. 1961 *Physical Review* **124**, 246
6. Arima, A. & Iachello, F. 1976 *Annals of Physics* **99**, 253-317
7. Gambhir, Y. K., Ring, P. & Schuck, P. 1983 *Physical Review Letters* **51**, 1235-1238
8. Pines, D. & Alpar, M. A. 1985 *Nature* **316**, 27-32
9. Baym, G., Pethick, C. & Pines, D. 1969 *Nature* **224**, 673-674
10. Blatt J. M. 1964 *Theory of Superconductivity* (Academic Press Inc, New York) for shape resonances in a metallic membrane see 362 pp and 215 pp
11. Innocenti, D. *et al.* 2010 *Physical Review B* **82**, 184528
12. Perali, A., Bianconi, A., Lanzara, A. & Saini, N. L 1996 *Solid State Communications* **100**, 181-186
13. Bianconi A, Valletta A, Perali A, Saini N L, 1998 *Physica C* **296**, 269-280

14. Lifshitz I. M. 1960. *Sov. Phys. JEPT* **11** 1130.
15. Bianconi, A. 2005 *Journal of Superconductivity* **18**, 625-636
16. Leggett AJ 1980 *Modern Trends in the Theory of Condensed Matter* edited by A. Pekalski and R. Przystawa, Lecture Notes in Physics Vol. 115 (Springer-Verlag, Berlin). (see p. 13); Leggett A 1989 in *The New Physics* P. Davies ed., Cambridge University Press pp. 268-288
17. Perali, A., Innocenti, D., Valletta, A. & Bianconi, 2012 A. Superconductor Science and Technology 25, 124002.
18. Fano U. , 1935 Il Nuovo Cimento (1924-1942) **12**, 154
19. Fano U, 1961 *Phys. Rev.* **124**, 1866
20. Bianconi A., 2003 Ugo Fano and shape resonances. AIP Conference Proceedings 652,13-18. doi:http://dx.doi.org/10.1063/1.1536357
21. Vittorini-Orgeas, A. and Bianconi, 2009 A. Journal of Superconductivity and Novel Magnetism 22, 215
22. Bianconi A. 1993 "High $T_c$ superconductors made by metal heterostructures at the atomic limit" European Patent No. 0733271 published in European Patent. Bulletin 98/22, May 27 1998 (priority date 7 Dec 1993)
23. Bianconi, A. *et al.* 2001. Journal of Physics: Condensed Matter **13**, 7383-7390
24. Di Castro, D. *et al.* 2002 EPL (Europhysics Letters) **58**, 278-284
25. Bussmann-Holder, A. & Bianconi, A., 2003 Physical Review B **67**, 132509
26. Laliberte F., Chang J., Doiron-Leyraud N., Hassinger E., Daou R., Rondeau M., Ramshaw B. J., Liang R., Bonn D. A., Hardy W. N., Pyon S., Takayama T., Takagi H., Sheikin I., Malone L., Proust C., Behnia K., and Taillefer 2011 L. *Nature Communications* 2, 432.
27. Liu C et al. 2011 Phys. Rev. B **84,** R020509
28. Ideta S et al 2013. Phys. Rev. Lett. 110, 107007
29. Innocenti D, Caprara S, Poccia N, Ricci A, Valletta A, and Bianconi A, 2011 Supercond. Sci. Technol. **24**, 015012
30. A. Bianconi, M. Missori, H. Oyanagi, H. Yamaguchi, Y. Nishiara, and S. Della Longa, 1995 EPL (Europhysics Letters) **31**, 411-415
31. Bianconi, A. & Missori, M 1994 Solid State Communications **91**, 287-293
32. Bianconi, A. Doniach, S. and Lublin, D. 1978 Chemical Physics Letters 59, 121
33. Garcia J, Bianconi, A, Benfatto M., and Natoli C. R., 1986. Le Journal de Physique Colloques 47, C8-49. DOI: 10.1051/jphyscol:1986807
34. Della Longa, S., Soldatov, A., Pompa & Bianconi, A. 1995 Computational Materials Science **4**, 199-210 .
35. Bianconi A, Saini N L, Lanzara A, Missori M, Rossetti T, Oyanagi H, Yamaguchi H, Oka K and Ito T 1996 *Phys. Rev. Lett.* **76,** 3412.
36. Bianconi, A. et al. 1996 . Physical Review B 54, 4310
37. Bianconi, A. Li C., Pompa M., Congiu-Castellano A., Udron D., Flank A. M., and Lagarde P., 1991 Physical Review. B, Condensed matter 44, 10126
38. Poccia N., Ricci A. Campi G. , Fratini M. , Puri A. , Di Gioacchino D. , Marcelli A. , Reynolds M. , Burghammer M. , Saini N. L., et al. 2012, Proceedings of the National Academy of Sciences 109, 15685 (
39. Poccia N, Fratini M, Ricci A, Campi G, Barba L, Vittorini-Orgeas A, Bianconi G, Aeppli G, Bianconi A, 2011 *Nat Mater* **10**, 733-736
40. Campi, G. *et al.* 2013 *Physical Review B* **87**, 014517
41. Jarlborg, T. & Bianconi, A. 2013 Physical Review B 87, 054514
42. Bianconi A, 2000. International Journal of Modern Physics B 14, 3289.
43. Bianconi A. , Di Castro D. , Bianconi G. , Pifferi A. , Saini N. L., Chou F. C., Johnston D. C., and Colapietro, M. 2000, Physica C: Superconductivity **341-348**, 1719 *Proc. International Conference on Materials and Mechanism of Superconductivity and High Temperature Superconductors (M2S-HTSC-VI), Feb. 20-25, 2000, Huston*
44. Agrestini S. , Metallo C. , Filippi M. , Simonelli L. , Campi G. , Sanipoli C. , Liarokapis E. , De Negri S. , Giovannini M. , Saccone A. , et al., 2004 Physical Review B **70**, 134514
45. Simonelli L. , Palmisano V. , Fratini M. , Filippi M. , Parisiades P. , Lampakis D. , Liarokapis E. , and Bianconi A. , 2009 Physical Review B **80**, 014520
46. Caivano R. , Fratini M. , Poccia N. , Ricci A. , Puri A. , Ren Z.-A.,. Dong X.-L, Yang J. , Lu W. , Zhao Z.-X., et al., 2009 Superconductor Science and Technology **22**, 014004
47. Ricci, A., et al. *Intrinsic phase separation in superconducting K0.8Fe1.6Se2 (Tc=31.8K) single crystals. Superconductor Science and Technology* **24**, 082002
48. Ricci, A., Poccia, N., Campi, G., Joseph, B., Arrighetti, G., Barba, L., Reynolds, M., Burghammer, M., Takeya, H., Mizuguchi, Y., Takano, Y., Colapietro, M., Saini, N. L., & Bianconi, A. 2011 Phys. Rev. B, 84, 060511
49. Kresin, V. Z. Ovchinnikov, Y. N. and. Wolf, S. A. 2006 : Physics Reports **431**, 231
50. Bishop, A. R. 2008 *High $T_c$ oxides: a collusion of spin, charge and lattice.* Journal of Physics: Conference Series 108, 012027 . doi:10.1088/1742-6596/108/1/012027
51. Kugel, K. I., Rakhmanov, A. L., Sboychakov, A. et al. A. 2008. *Physical Review B 78, 165124.*

52. Ovchinnikov S G, Korshunov M. M., Shneyder E. I. 2009. Journal of Experimental and Theoretical Physics 109, 775.
53. Ovchinnikov, S. G. Korshunov, M. M. Makarov, I. A.. Shneyder E. I 2013 Journal of Superconductivity and Novel Magnetism doi:10.1007/s10948-013-2144-1
54. Balakirev F. F., Betts J. B. , Migliori A., Tsukada I., Ando Y., and G. S. Boebinger 2009 Phys. Rev. Lett. 102, 017004.
55. Bianconi G, 2012 Physical Review E, 85, 061113
56. Bianconi G, 2012.*Journal of Statistical Mechanics: Theory and Experiment* **2012**, P07021 ;
57. Bianconi, 2013 G.. *EPL (Europhysics Letters)* **101**, 26003
58. Halu A, Ferretti L, Vezzani A and Bianconi G 2012 *EPL (Europhysics Letters)* **99**,18001
59. García-García A.M. Thermal 2013 *Amplitude Fluctuations in Arrays of Small Josephson Junctions* Journal of Superconductivity and Novel Magnetism doi: 10.1007/s10948-013-2162-z
60. *Stripes and related phenomena*, edited by Antonio Bianconi and Naurang L. Saini Published in 2000, Kluwer Academic/Plenum Publishers (New York)
61. *Symmetry and heterogeneity in high temperature superconductors* edited by A. Bianconi, Springer New York,Berlin (2006), Proc. Advanced Study Research Workshop on Symmetry and Heterogeneity in High Temperature Superconductors, North Atlantic Treaty Organization (NATO) Science Series II: Mathematics, Physics and Chemistry)
62. Boninsegni M. and Prokof'ev, N. V. Reviews of Modern Physics **84**, 759 (2012),